\RequirePackage{fix-cm}
\documentclass[twocolumn]{svjour3}
\smartqed

\usepackage{graphicx}
\usepackage{picinpar}
\usepackage[utf8]{inputenc}
\usepackage[hidelinks]{hyperref}
\usepackage{microtype}

\journalname{}

\begin{document}
\title{
Managed Forgetting to Support Information Management and Knowledge Work
%
\thanks{This work was funded by the Deutsche Forschungsgemeinschaft (DFG, German Research Foundation) -- DE 420/19-1.}
}

\author{
  Christian Jilek\textsuperscript{1,2} \and
  Yannick Runge\textsuperscript{3} \and
  Claudia Niederée\textsuperscript{4} \and
  Heiko Maus\textsuperscript{1} \and
  Tobias Tempel\textsuperscript{5} \and
  Andreas Dengel\textsuperscript{1,2} \and
  Christian Frings\textsuperscript{3}
}
\authorrunning{Christian Jilek et al.}

\institute{
  \textsuperscript{1}
  Smart Data \& Knowledge Services Department, DFKI GmbH,
  Kaiserslautern, Germany\\
  \email{
    \texttt{christian.jilek@dfki.de},
    \texttt{heiko.maus@dfki.de},
    \texttt{andreas.dengel@dfki.de}}
  \\[0.2cm]
  \textsuperscript{2}
  Knowledge-Based Systems Group,
  Department of Computer Science, TU Kaiserslautern,
  Kaiserslautern, Germany
  \\[0.2cm]
  \textsuperscript{3}
  Department of General Psychology and Methodology,
  University of Trier, Trier, Germany\\
  \email{
    \texttt{runge@uni-trier.de},
    \texttt{chfrings@uni-trier.de}}
  \\[0.2cm]
  \textsuperscript{4}
  L3S Research Center,
  Hannover, Germany\\
  \email{\texttt{niederee@l3s.de}}
  \\[0.2cm]
  \textsuperscript{5}
  Special Education Department,
  Ludwigsburg University of Education,
  Ludwigsburg, Germany\\
  \email{\texttt{tobias.tempel@ph-ludwigsburg.de}}
}

\date{}

\maketitle

\begin{abstract}
Trends like digital transformation even intensify the already overwhelming mass of information knowledge workers face in their daily life.
To counter this, we have been investigating knowledge work and information management support measures inspired by human forgetting.
In this paper, we give an overview of solutions we have found during the last five years as well as challenges that still need to be tackled.
Additionally, we share experiences gained with the prototype of a first forgetful information system used 24/7 in our daily work for the last three years.
We also address the untapped potential of more explicated user context as well as features inspired by Memory Inhibition, which is our current focus of research.

%
\keywords{
  (Intentional) Forgetting \and
  Information management and knowledge work support \and
  Semantic desktop \and
  User context \and
  Self-organizing systems
}
\end{abstract}


\section{Introduction}
The ever increasing flood of information knowledge workers face in their daily lives even intensifies by technological trends like digital transformation.
Thus, their usual multi-tasking craziness \cite{GonzalezMark04}, constantly switching from one context to another, each being associated with different tasks, documents, mails, etc., gets even worse.
As a typical consequence, their personal information space, such as file/mail/bookmark folders, is cluttered with information that has become irrelevant.
Thus, finding important information gets harder and much of previously gained knowledge is practically lost.

To address these problems, we have been investigating solutions inspired by human forgetting since 2013, starting with the EU-project
\textit{ForgetIT}\footnote{2013--2016, \url{www.forgetit-project.eu}}
and continuing in the
\textit{Managed Forgetting} project\footnote{2016--2019, \url{www.spp1921.de/projekte/p4.html.de}},
which is part of the recent priority program on ``Intentional Forgetting in Organizations''  by the German Research Foundation (DFG).
Together with other teams of this program we already presented an overview on perspectives and challenges of intentional forgetting in artificial intelligence systems in general \cite{TimmStaabSiebers+2018}.

In this paper, we complement that survey having a particular focus on knowledge work and information management support.
First, we give an overview of solutions we already found in the two aforementioned projects.
We especially share experience gained with the prototype of a first forgetful information system (FIS) that we have been using 24/7 in our daily work for the last three years (Section 2).
Additionally, we point out which challenges still need to be tackled, give insights on how we intend to address them, or present first solutions or prototypes that are still under development (Section 3).
Section 4 concludes this paper and gives an outlook on planned next steps.

\section{Towards Forgetful Information Systems in Practice}
In the \textit{ForgetIT} and \textit{Managed Forgetting} projects, we investigated knowledge work and information management support measures inspired by human forgetting.
Especially in the second project, our investigation is positioned in the context of a grass-roots Organizational Memory (OM), which relies on the principles of decentralization and self-organization:
Effective, dynamic and tailored knowledge management is achieved by knowledge-based assistance and knowledge acquisition in daily activities of knowledge workers which in turn also shapes the captured and represented knowledge.
Following the \textit{eat-your-own-dogfood} credo, we extended our OM system, which we have been using in daily work for over seven years now, with forgetting mechanisms, in use now for the last three years.
Before presenting its details, also serving as solutions to the aforementioned problems, we will first give an introduction into the terminological and technical background.

\subsection{Managed Forgetting}
As an extension to the binary keep-or-delete paradigm, we understand \textit{Managed Forgetting} (MF) \cite{KanhabuaNS13,NiedereeKanhabuaGallo+15,NiedereeMezarisMaus+2018} as an escalating set of measures: from temporal hiding, to condensation, to adaptive synchronization, archiving and deletion.
It is a form of intentional forgetting that is completely based on observed evidences:
the system learns what to forget and what to focus on in a self-organizing and decentralized way.

As a key concept for realizing this form of MF we have presented \textit{Memory Buoyancy} (MB) \cite{NiedereeKanhabuaGallo+15,NiedereeKanhabuaTran+2018}, which is intended to represent an information item's current value for the user.
It follows the metaphor that items which start to lose relevance for the user ``sink away'', while those that are important are pushed closer to the user by their higher buoyancy.

\textit{Information Value Assessment} (IVA) for deciding about the current importance of an information item~is core for dynamically determining its MB value.
IVA~in the context of MF has been investigated for \mbox{individual} types of resources such as photos \cite{CeroniSolachidisNiederee+2015} as well as in broader terms for the resources on a user's desktop \cite{TranSchwarzNiederee+2016,MausJilekSchwarz2018}.

Like stated before, this form of MF requires capturing and interpreting evidences in order to work.
We chose the \textit{Semantic Desktop}, which will be discussed in the following, to serve this purpose.

\subsection{The Semantic Desktop as an ecosystem for Managed Forgetting}
\label{sec:SemDesk}
\paragraph{The Semantic Desktop \& PIMO.}
The \textit{Semantic Desktop}
(SemDesk) \cite{sauermann2005semdesk} is especially intended to capture knowledge that emerges from individuals and then spreads into groups like project teams.
SemDesk brings \textit{Semantic Web}\footnote{\url{www.w3.org/standards/semanticweb}} technology to users' computing devices using a knowledge representation, i.e. giving resources unique identifiers (URIs) and allowing to make statements about them, e.g. using RDF\footnote{\url{www.w3.org/RDF}}, resulting in a semantic graph.
Information items (files, mails, contacts, events, topics, \ldots) that are separated on the computer (file system, mail client, web browser, \ldots) but are related to each other in a person's mind, can thus be semantically represented and interlinked in a machine understandable way.
As soon as such an item is semantically represented, it is called a ``thing'', which describes the item uniquely as an URI complemented by further statements like its type or a reference to the originating resource such as an URL or message-id of an e-mail.

Capturing a user's mental model as accurate as possible is done in a \textit{Personal Information Model} (PIMO) \cite{SauermannVanElstDengel2007}, which serves as the basis for knowledge representation in SemDesk.
Shared parts of multiple PIMOs result in a \textit{Group Information Model} (GIMO) forming the basis for an OM.

\begin{figure}
  \includegraphics[width=1\columnwidth]{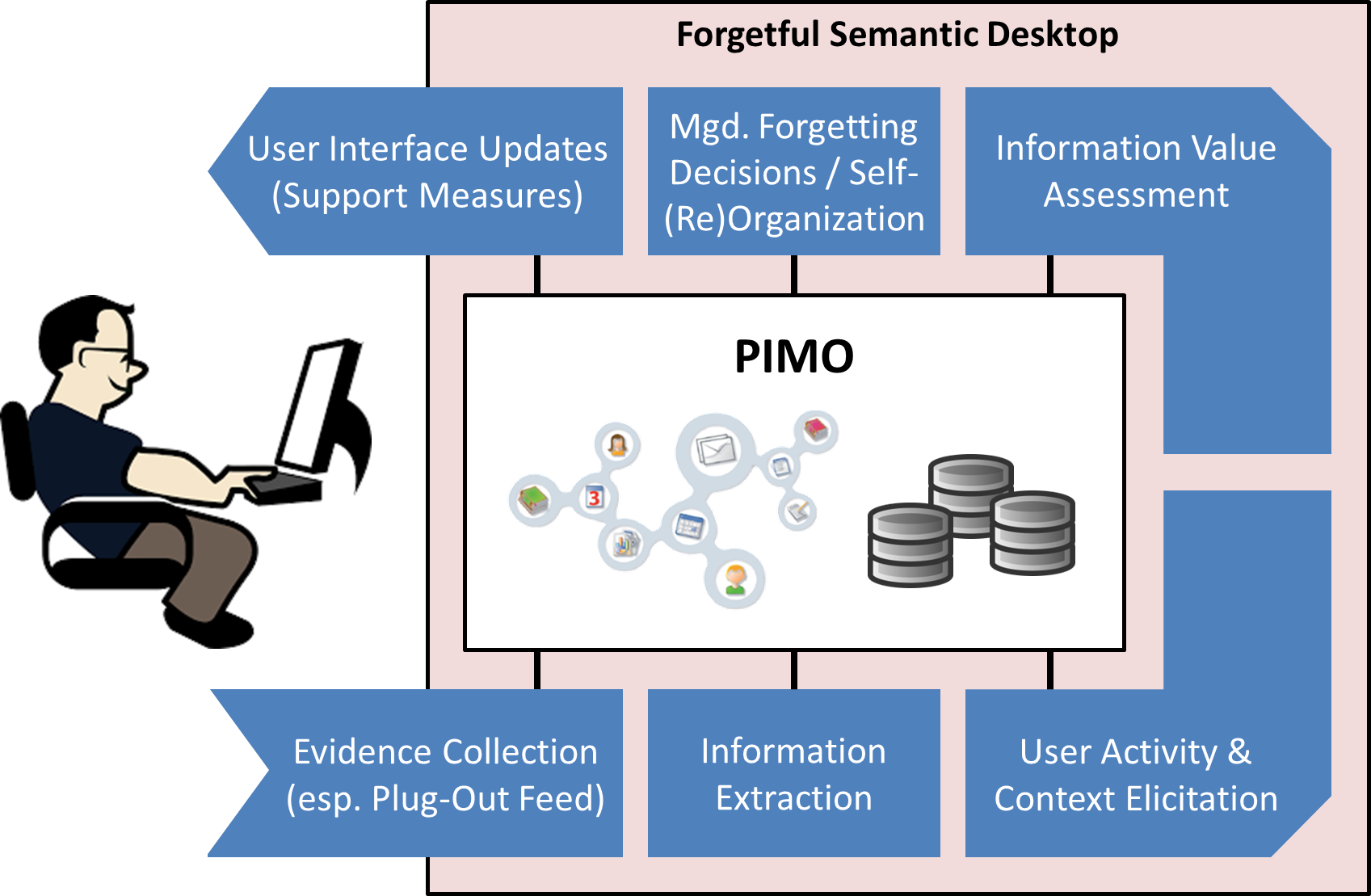}
  \caption{Evidence collection to user support measures cycle of a Forgetful Semantic Desktop}
  \label{fig:ForgetfulSemDesk}
\end{figure}

\paragraph{From Evidence Collection to User Support Measures.}
Concerning SemDesk applications, two categories, newly created semantic ones and plug-ins to enhance traditional, non-semantic ones, could be observed so~far~\cite{DraganDecker2012}.
We recently presented our idea of \textit{Plug-Outs} \cite{JilekSchroederSchwarz+2018}, \textit{headless plug-ins} often just having the rudimentary functionality of \textit{sending out} in-app events to the SemDesk.
Complementing these plug-outs with the transparent integration of SemDesk using standard protocols and a sidebar for advanced features \cite{JilekSchroederSchwarz+2018}, we get an environment capable of capturing rich contextual evidences in two ways: implicitly (plug-out and protocol information) as well as explicitly (sidebar usage).
These evidences are then processed further, especially in terms of information extraction \cite{JilekSchroederNovik+2018}, to elicit the user's current activity and context.
This results in the respective stimulation of the user's PIMO and appropriate MF measures.
Currently, we observe the file system, web browsers and email clients.
Further tools for process and application observation, especially also using accessibility interfaces \cite{HertlingSchroederJilek+2017}, are under development.
A comparative, yet incomplete literature overview of user activity tracking endeavors can be found in \cite{Schmidt2013PhD}.
In summary, the whole cycle from evidence collection to user support measures of a \textit{Forgetful} Semantic Desktop is depicted in Figure~\ref{fig:ForgetfulSemDesk}.

\subsection{First prototype running 24/7 in practice}
\label{sec:Prototype}
In \textit{ForgetIT}, we extended our SemDesk prototype \cite{MausSchwarzDengel2013}, that we were already using 24/7 in daily work, with MF features.
These new forgetting capabilities were thus directly embedded in daily activities, too, enabling us to continuously test and optimize them in real-world scenarios as well as understand implications and challenges of a forgetful information system.

To illustrate the challenge for the SemDesk usage, let us look at one user there.
As of July 2018, the semantic graph of his PIMO consists of more than 18.000 things, i.e., topics, tasks, events, persons, organizations, documents, web pages, images, notes, \ldots, of which 2.000 are private ones and the rest either shared by him or other users forming the group's GIMO.
There are not only noteworthy or timeless things such as scientific papers or project proposals among them, but also various ephemeral things.
For instance, more than 1.000 tasks from over 7 years of daily usage leave their electronic footprints (including connected resources, topics, or notes) in the PIMO.
Now, these once relevant tasks bear the potential to serve as a task journal -- if not even as source for know-how reuse -- but also to congest the semantic graph and search results.

Therefore, it is evident that the user would face an information overflow, if each and every thing would be treated with the same importance as currently required information or important one from the past;
PIMO's usefulness would thus be endangered.
But requiring to delete presumably outdated information counteracts the philosophy of our evolving knowledge management system due to the potentials for future situations of the individual and the group.

\paragraph{Memory Buoyancy Calculation.}
The MB calculation in our SemDesk evolved from the insights of the approach in \cite{TranSchwarzNiederee+2016} and finally follows the design principles presented in \cite{MausJilekSchwarz2018}.
The basic ones are inspired by human brain activity  applied to the user's mental model as represented in a semantic graph and discussions with the team of Prof. Logie (Psychology, University of Edinburgh) who presented their insights in \cite{LogieWoltersNiven2018} (an interdisciplinary approach is also taken in the \textit{Managed Forgetting} project and is subject of Section~\ref{sec:potential}).
Therefore, the MB value drops over time for things that are not stimulated (first a steep decline then a long-tail of slow decline) whereas the value increases for things and their associations that are stimulated.
To reflect learning effects, the MB value decreases slower for things that are repeatedly stimulated over time.
The stimulations are based on evidences derived from user actions (such as view, create, or modify) involving resources represented as things in the semantic graph.

The intensity of stimulating a thing and the outreach along the sub-graph it spans is determined by applying a dedicated spreading activation algorithm \cite{Crestani1997} using parameters such as types of the things, connecting predicates as well as numbers of connections, and several heuristics.
This leads to effects like things such as topics, which are not directly accessed but are connected to resources currently in use, are raised in their MB value, then forming hotspots in the semantic graph resembling the user's current mindset (of those items represented in the~PIMO).

Apart from tuning the spreading algorithm, most influential on the IVA are the various heuristics applied.
These cover assumptions as well as intended beneficial effects for the knowledge work scenario.
For instance, applying specific decay curves for dedicated types, e.g. enforcing e-mails to decay faster than presentations, is based on the observation that inherently \mbox{e-mails} are more ephemeral whereas presentations might imply more longevity.
Further, upcoming events and connected things are stimulated when approaching their date and vice versa, decaying faster after the event (if not stimulated again).
Finished tasks do not get external stimulations any more.
This results in MB values for all things in a PIMO, which then can be used for various forgetting strategies outlined in the following.

\paragraph{Temporal Hiding.}
A first escalation step in our MF strategy is hiding things which are below a certain threshold in different places where users might be confronted with an overwhelming amount of information items from the semantic graph. 
Therefore, while browsing the PIMO, in which a thing is usually represented on a single (HTML) page with details (such as start and end date of an event) and all its relations to other things, connected things below a specific MB threshold are hidden from direct view of the user.
This threshold is lower on desktop devices and higher on mobile devices to reduce cognitive load while being mobile, i.e., showing less presumably irrelevant information.
A button ``show forgotten'' allows to show the yet hidden things as well as to manipulate the threshold for viewing.

A further experimental feature is hiding things in a search result list with low MB value which simply hides those things to be forgotten from the result list.
The search page has a threshold slider to lower or raise the currently set threshold for the specific search allowing to blend in those forgotten things in the position of their original ranking.

Considering snippets used for explaining the content of result list entries (of a search or a proactive information delivery), the MB value is also used to choose a set of things to be shown in the snippet.
Here, instead of deciding to hide, the sub-set of high buoyant things out of all annotations of a thing are selected and shown in the result entry as snippets. The assumption is that showing all annotations of a document will overwhelm the user, instead selecting the ones which are high-buoyant will allow to give a clue to quickly grasp the relevancy for the user's mindset.

\paragraph{Adaptive Synchronization.}
Next step in our escalating MF strategy is to move files to be forgotten from the desktop to the cloud and finally to an archive.
This still keeps the semantic representation available, only the place of the actual file changes.
If access is required, it can be automatically drawn in again.
This is an experimental feature embedded in the PIMO cloud synchronization service running on a users' desktop.
Current implementation proposes the user a list of files (which are things) that can be forgotten on the computer, selected files are created as cloud-files (if not already there) and then removed locally.
Moreover, an extension enables an unsupervised adaptive synchronization of files to a local storage if their MB is above or below a certain threshold (which again can be different depending on the device type).
This is useful for files which, e.g., originate from other users and are not yet available on the device.
The PIMO's user interface then allows to open the local version directly instead of downloading it first.
This leads to a set of documents on the device which are relevant for the user's current mindset (and are not elsewhere on the device anyway).

\paragraph{Condensation.}
Further action can be undertaken if a whole region of the semantic graph has a low MB such as a long-finished task or project.
Then it is possible to condense this region (consisting of things and connected resources) and just leave a representation of this region for the user.
Our PIMO Diary \cite{JilekMausSchwarz+2015,JilekSchwarzMaus+2016} uses condensation to on demand generate condensed representations for the user's electronic footprints within a specific period of time.
The condensed representation is shown to the user including associations to the contained things.
This bears the potential of forgetting the originating things and resources (e.g., by moving them to an archive) while keeping references in the condensation.
This explicit removal is not deployed on our PIMO, however, once computed condensations are kept.

\paragraph{Lessons Learned.}
The MB calculation is in use for~over 3 years.
Our experience so far is that things are really gradually fading out if their relevancy decreases. And vice versa, related things raise in their MB if they are connected to user activities although they are not directly accessed, thus forming a user's recent mindset.

In contrast, there are also drawbacks: the naive approach for hiding in search has been dropped. An often observed behavior was that users moved the slider to change the threshold to zero (i.e., show everything) if the results were not satisfying, instead of modifying the query first; whereas with lots of results, the slider was ignored. This implies that there is still not enough trust in MF if the results are not as expected.

If things are raised in their MB although they are not explicitly accessed heavily depends  on their connectivity in the graph.
Thus, isolated areas might drop although the relevancy is still given.
Here, more automated interconnection is required and we see that if they belong to some context, things are dragged along with the rest.

Likewise, raising the MB for a whole area takes some time if the user jumps into a previously neglected area.
Several actions are required to differentiate between visiting by chance or really working in that area again.
This also implies to consider contexts which can be revisited after a long time but the user would expect everything in place in analogy to the brain which is able to quickly reconstruct a scenery.

Up to now, we only discussed the forgetting aspect in our scenario. But from the knowledge management viewpoint also the aspect of long-term information value plays an important role.
In \cite{MausJilekSchwarz2018} we also considered a
``preservation value'' as an orthogonal view on things
for their long-term importance although they might only have had a short-term relevancy for the user.\\

From these experiences and projections with other research threads, we identified challenges which will be addressed in Section~\ref{sec:Challenges}.

\subsection{Untapped potential of explicated user context and features inspired by Memory Inhibition}
\label{sec:potential}
\paragraph{Explicated User Context.}
From the experiences presented in the last section, we learned that a greater focus on user context can further improve solutions found so far.
We especially assume that users are aware of the concept of context and what their current context is (at least most of the time) \cite{GomezPerez2009}.
In \cite{JilekSchroederSchwarz+2018} we presented a first SemDesk prototype\footnote{demo video at \url{https://pimo.opendfki.de/cSpaces/}}, that has context as an explicit element users can work in and interact with.
We will go into details in Section \ref{sec:Context}.
Additionally, our current information value assessment can be improved by introducing context-sensitive MB values (see \ref{sec:AdvMB}).

\paragraph{Memory Inhibition.}
In cognitive psychology, the term \textit{Memory Inhibition} describes the temporal suppression of currently irrelevant or misleading information in order to facilitate processing of relevant information \cite{levy2002inhibitory}.
Cognitive psychology experiments like \cite{TempelFrings2016} revealed that intentionally forgetting about one task can enhance subsequent cognitive performances like encoding and recall of word material.
A prominent explanation for these benefits is memory inhibition of intentionally forgotten information \cite{Bjork1989}.
Inhibition can also help to efficiently switch contexts, by mentally segregating irrelevant, inhibited information from currently relevant information \cite{StormStone2015}.
We intend to transfer these results to user contexts in knowledge work assuming that allowing users to intentionally forget about their recently irrelevant contexts increases their performance on the current one.

To allow users to intentionally forget, we will implement features inspired by Memory Inhibition, a concept, to our best knowledge, so far -- if at all -- only implicitly used in computer science without explicitly calling it that way.
In \cite{TempelNiedereeJilek+2018}, we give an overview of Memory Inhibition in cognitive and computer science and especially its potential for the latter.

To get an impression of how one could define inhibition in computer science, consider the following example:
information items associated with different contexts are activated by Spreading Activation (SA) \cite{Crestani1997}.
Then, items of those contexts, that have been activated but are irrelevant for a target context, will be suppressed (inhibited).
So, there is an additional differentiation mechanism for items correctly activated by classic SA with respect to a given target context.
Thus, inhibition would be implemented to reduce or overcome interference due to information that is irrelevant for a particular target context.
Importantly, if contexts switch again, inhibition of previously irrelevant items is released -- thus inhibiting items always means making items temporarily unavailable.

\section{Challenges} \label{sec:Challenges}
The last section already gave insights into which challenges we had to tackle to get a beneficial forgetful information system, e.g. to establish continuous user activity tracking or memory buoyancy calculation.
In this section, we will address open challenges or ones that we have only solved partly so far.
For the latter, we will also give insights on how we intend to solve them.
Although we focus on a system to support information management and knowledge work, some of the solutions may also be applicable to FIS in other domains.

\subsection{Capturing and efficiently storing metadata especially contextual information}
\label{sec:Context}
\begin{figure}
  \center
  \includegraphics[width=1\columnwidth]{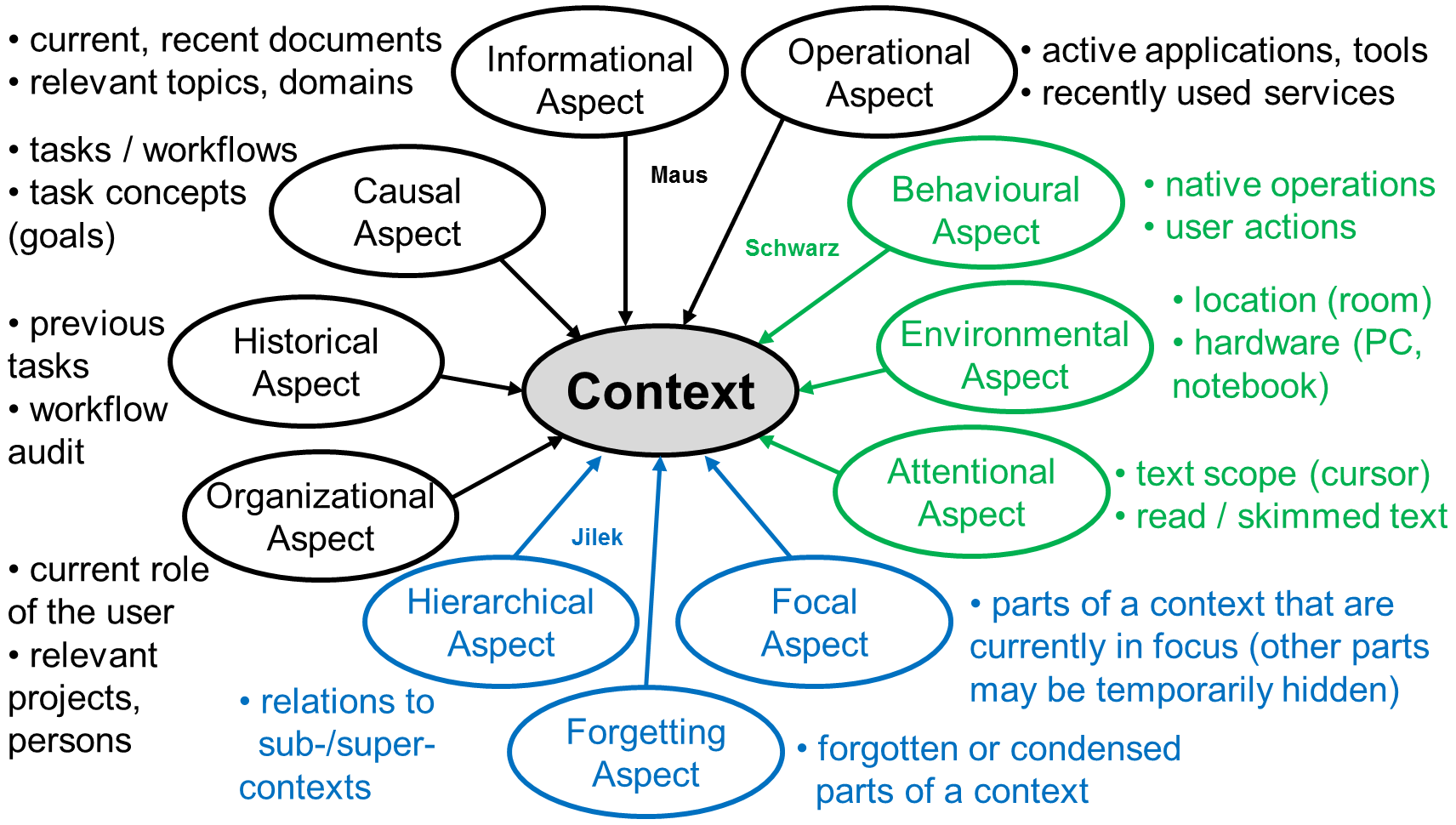}
  \caption{By adding hierarchical, forgetting and focal aspects (blue), we extend the context model by Schwarz \cite{Schwarz2005}, which itself is an extension (green) of the one by Maus \cite{Maus2001} (black).}
  \label{fig:ContextModel}
\end{figure}

\paragraph{Context.}
We learned from cognitive psychology (see \ref{sec:potential}) that our previous research on context (e.g. \cite{Schwarz2010,MausSchwarzHaas+2011,JilekMausSchwarz+2015}) can be beneficial for MF:
an information item can be very important in one context while being totally irrelevant in another.
So, in order to finally decide about an item's relevancy and thus provide beneficial MF measures, we have to take its associated contexts into account.
The context model we use is depicted in Figure \ref{fig:ContextModel}.
It is an extension of \cite{Schwarz2005}, which itself is an extension of \cite{Maus2001}.
For our use case we additionally added the following aspects:
\textit{forgetting} (which parts of a context have been forgotten or condensed),
\textit{focus} (which parts of a context are currently in focus; other parts may be temporarily hidden, for example), and
\textit{hierarchy} (sub-/super-contexts).

\paragraph{Capturing Context.}
In order to take contexts into account, we have to capture them first, which implies the usage of sensors.
Currently, we completely focus on virtual sensors \cite{PereraZaslavskyChristen+2014} as realized by our plug-outs (see \ref{sec:SemDesk}), for example.
Since the ultimate answer which items belong to which contexts is only available in a person's mind, there will probably always be a certain sensor and interpretation gap.
Thus, we can only try to minimize it and approximate user contexts as well as possible.
The aforementioned idea of treating contexts as explicit elements that users can work in and interact with (see \ref{sec:potential}) also helps us in this regard:
users can casually help in modeling their world by selecting their current context, working with it (e.g., to accomplish a task), adding or removing things, switching it, etc.
We will come back to this aspect in Section \ref{sec:Incentives}.

\paragraph{Storing Context.}
Once we are able to capture contexts, we also have to store them efficiently together with other (meta)data managed by our system, especially the semantic graph (PIMO).
Since several support decisions need to be made in less than a second in order not to harm user experience (system response time), we especially need data structures capable for real-time processing on usual computing devices.
The same is true for the used information extraction methods: they should be able to operate in (near) real-time.

\paragraph{Privacy Issues.}
Another aspect we have to take into account are privacy issues that arise immediately when dealing with any kind of user activity tracking.
Since we are capturing possibly very sensitive data, we have to take measures to protect users' privacy.
Possible solutions are, for example, allowing to (temporarily) disable the observation, only store sensitive data on the user's local device (no server/cloud sync) or only stimulate the semantic graph (``activation'' of respective parts) without storing any details permanently.

\subsection{Continuous, context-sensitive information value assessment}
\label{sec:AdvMB}
We have presented our current version of information value assessment resulting in different MB values in Section \ref{sec:Prototype}.
Doing IVA continuously is already a challenge, since each click of a user may alter the MB of possibly a lot of things (depending on semantic network connectivity).
One problem of our current solution is that there is only one MB value for a resource for each user, without taking contexts into account.
But, as stated before, an item's relevancy can strongly vary from one context to another.
Thus, we are in the process of advancing our MB calculation to additionally take contexts into account \cite{JilekChwalekSchwarz+2018}.
If a resource is not associated with a context, or not accessible by the user, its MB value for this user and context is zero.
In cases, in which we know that a certain activity (tagging a website, opening a document, etc.) was performed having selected a certain context (e.g. \textit{ForgetIT project proposal}), we thus only have to update a relatively small part of the semantic network compared to the original version, which also leads to performance gains.
We call this the \textit{local MB}, since it is only calculated for a certain context.
Nevertheless, we also keep a context-free MB value, the \textit{global MB}, summarizing all non-zero local MB values, thus providing an overall relevancy information of a resource for a certain user.
Especially with regard to OM, we additionally introduce a \textit{group MB} that summarizes the global MB values of different users.
More details and a first prototypical implementation are presented in \cite{JilekChwalekSchwarz+2018}.
\subsection{Improve user interfaces and support features to enable cognitive offloading}
Have you ever turned your head to read a tilted headline more easily?
If this is the case, you have been performing a form of \textit{Cognitive Offloading} \cite{RiskoGilbert2016}.
Every strategic use of physical actions like tilting your head or using a calculator to reduce cognitive demands is defined as \textit{Cognitive Offloading}.
Only two decades ago, people still learned phone numbers by heart in order to be able to call somebody.
Today, you just select the person's name in your (smart)phone and a connection will be established (assuming that you once have added that person's data to your contact list).
That way, we are using several devices as our extended external memory store.
Recent research in cognitive psychology shows that such memory offload can have comparable benefits for subsequent cognitive performance as discussed for intentional forgetting in \ref{sec:potential} \cite{StormStone2015,RungeFringsTempel2018}.
As shown in \cite{RungeFringsTempel2018}, the possibility to store information externally can even be seen as an implicit cue to intentionally forget the offloaded information as long as you can rely on the device to continually store it \cite{StormStone2015}.

In our scenario, think of contexts that ``tell'' the users what they have done the last time when they were working in/with them:
what documents have been read or written, what are open tasks to be performed, etc.
Such implementation allows the user to rely on the SemDesk as their external memory for all reached work progress.
The intentional forgetting of externally stored work progress can then benefit subsequent tasks and ease the switch from one context to another.
Due~to our transparent integration \cite{JilekSchroederSchwarz+2018}, these contexts are also available as folders in the file system.
Thus, the metaphor of folders presenting a user their ``golden thread'' is close to realization.
Our ultimate goal would be to also bring back all applications in the exact state they were in when they were last opened in that certain context.
But this is a task, which is hard to perform due to missing interfaces for application resume.
In general, we have the challenge of creating user interfaces fitting well with our MF capabilities, so that users are actually able to cognitively offload.
If they still have to keep rather unimportant or too many things in mind or bother how to store things in a way that they will find them later, cognitive offloading is not possible.
It is the system that should note what they did last, that the link to a certain website is stored for next time, that certain reminders or documents come up as soon as they become relevant (again), etc.


\subsection{Incentives to maximize users' willingness to contribute}
\label{sec:Incentives}
In Section \ref{sec:Context}, we gave reasons why we will not be able to have a fully automated system.
We will rely on users helping to make aspects of their mental model explicit in their PIMO.
Therefore, we should provide incentives that make them actually willing to do it, e.g. if they add a sent e-mail to a certain context, an incoming reply could automatically be associated with that context, too.
Further, the investments they have to spend should be as low as possible.
We thus have the challenge of designing interfaces and functionality so well, that a single click or drag operation can already mean a lot, for example.

This goes along the important aspect that users should immediately have (and see!) benefits from an action such as annotating a web page with a task (which in turn is an explicit ``modelling'' act in the semantic graph done by the user in the annotation sidebar).
Therefore, from our experience in knowledge work support, it is important to embed the support into daily work of the users and trying to create a context in which information need can be derived and required information be provided. Hence, it is important to find scenarios and use cases where this support is beneficial for users such as process work embedded in the e-mail client \cite{LampasonaRostaninMaus12} or a context space for solving tickets \cite{JilekSchwarzMaus+2016}.
Here, each action in that environment immediately leads to benefits for the user.

An inadequacy of investment and resulting benefit may lead to a vicious circle of knowledge management \cite{ProbstRaubRomhardt97}.
Our system should support the way towards a ``perfect model'' (i.e. user's PIMO and mental model are perfectly in sync), by allowing a sequence of tiny activities.
The SemDesk ecosystem already takes this direction by crawling information sources such as calendars, providing a sidebar allowing semantic bookmarking, writing semantic notes, or a task management.
For tasks such activities could be: create task, set deadlines, add notes, web or file links, etc.
Thus, the user can decide at any point whether they go another step or stop, whereas both, the system and the user, benefit from each additionally taken step.
Especially in the aforementioned multi-tasking craziness, in which users are under high pressure to continue their work and not spend time with seemingly unnecessary steps, they may thus easier regulate the amount of distraction they are currently willing to accept.
One of our hypotheses is, that even if the return on investment of a ``modelling activity'' is quite high, a too high corresponding user investment may prevent the activity from being performed by the user.
In contrast, having an interruptible sequence of tiny actions more likely leads to users doing at least some of the ``modelling steps''.

Other incentive measures, possibly well applicable in OM scenarios, are discussed in \cite{Lazaruk2012}, for example.

\subsection{Forgetful information systems need to be cautious and trustworthy}
In experiments, interviews and discussions conducted~in our forgetting-related projects, we observed the tendency of users, especially experts, to rather mistrust automated system decisions in fear of losing their stuff.
As a consequence, we strive to design our forgetful system to~be cautious, i.e. rather doing nothing than doing something wrong.
By acting this way, we seek to earn the users' trust; they need to be sure that nothing harmful (a data loss) will happen.
In particular, this means if collected and interpreted evidences do not justify to take a certain support action, the system will refrain from performing it.
Our colleagues \cite{SiebersGoebelNiessen+2017} go one step further and have their system not take any action without user confirmation, which is what we also did~in the past:
When the MB of files dropped below a certain threshold and the system selected them for being deleted on the user's current device, users were first asked for confirmation.
Additionally, since our system is still a research prototype, we did not delete any files completely so far -- there is always a backup in an archive.
Nevertheless, we abandoned asking for confirmation since it interrupted users and drawing their attention on actually forgotten regions resulting in the contrary effect of remembering, as well as counteracting our idea of \mbox{self-organization}.

In \cite{JilekSchroederSchwarz+2018}, we presented a first prototype of a self-reorganizing SemDesk based on MF features.
Folders can also be seen as contexts and the system is even able to infuse its managed contexts into the file system.
Following the principle of the aforementioned cautiousness, automatically reorganizing these contexts, e.g. merging or splitting them, should not lead to totally different paths that users have to follow in order to re-find desired information.
Since navigating folders (contexts) follows the human intuition of navigating a map \cite{BennBergmanGlazer+2015}, sudden, unexpected changes are potentially harmful here.
We could thus restrict our system to only merge parent and child contexts after some time has passed.
Greater modifications like merging contexts of similar but no directly related topic areas, for example, would only be allowed if a lot of time has passed and there was no new evidence indicating that one of the topics is still relevant.
Thus, the goal of automatically tidying things up for the user would justify such a merge of similar contexts.
\cite{BennBergmanGlazer+2015} also states that Personal Information Management (PIM) applications should take folders (contexts) as given and try to exploit and improve them rather than replace them, which is an advice we intend to follow as well as possible.

\subsection{How to gain trust in forgetful search and how to visualize the forgotten?}
One major question we still have to solve is how we achieve that users gain trust in forgetful search.
Consider the example of a user entering keywords into the search field of a forgetful system and no (or seemingly incomplete) results are shown.
Several questions could come to that user's mind:
\textit{Have I used the ``right'' keywords?
Have I really saved the things I am now looking for?
I'm sure I saved it, why doesn't it show up?}
The challenge of establishing trust in forgetful search is closely linked to the question of how to visualize what is actually forgotten (forgotten in the sense of MF, i.e. only a condensed version is still remaining or something is hidden by default due to low MB).
In the scenario just mentioned, we could inform the user that there is currently no search result in the ``active'' part of their data, but something in the forgotten area.
This could be accompanied by measures trying to visualize how search results belong to certain areas of the semantic network, e.g. as thematic clusters, as well as how much of the semantic graph has been covered by the current search result set.
Additionally, as done in the PIMO, if the user enters an exact match of a thing's label, e.g. the full name of a person or project, then it would be justified to directly show actually forgotten items, since the user seems to have remembered something they have not used for a long time.
In general, we have to find a balance between MF mechanisms that prevent users from being overwhelmed by the potentially high number of search results.
But on the other hand, users still have to find the things they are looking for, especially if they came back to something not accessed for a very long time (whereas ``accessed'' here especially means that a whole topic area of the semantic graph has not been stimulated for a long time -- accessing related topics would have raised the MB otherwise).

\subsection{How to evaluate forgetful information systems?}
All challenges mentioned so far share the problem of~how to evaluate their possible solutions.
Evaluating an FIS like ours is hard for several reasons.
First, since we support information management and knowledge work, users' views on their stuff are subjective \cite{Dengel2006}, which restricts evaluation scenarios.
Second, there is still no publicly available PIM dataset.
To our best knowledge, \cite{AbelaStaffHandschuh2015} is the most recent paper mentioning the plan to release a dataset ``in the near future'', which is already three years ago.
\cite{Gonccalves2011} even argues that if such a dataset was available, it would still lack the semantic information to really make use of the data (e.g. whether a term is the name of a project or whether a mentioned person is a co-worker or spouse, etc.).
Other approaches like \cite{KimCroft2009b} created pseudo desktop collections for their experiments (on information retrieval).
These collections neglect important sources like bookmarks or calendar events as well as structures like the file folder hierarchy, which also carry a lot of semantics.
Last not least, we have the additional aspect of forgetting, which makes us state the hypothesis that participants need to perform the evaluation of such systems using their own data.
How could people otherwise judge whether things were forgotten correctly if they never knew the data?

To solve this problem, we intend to semi-automatically bootstrap the semantic graph (PIMO) of a participant before starting the evaluation of forgetting capabilities.

\section{Conclusion \& Outlook}
In this paper, we gave an overview of information management and knowledge work support measures found in two forgetting-related projects.
In the first one, \textit{ForgetIT}, we enhanced our productively used, SD-based OM system with forgetting capabilities, thus having one of the first FIS used in practice.
From the beginning on, in 2013, we searched for solutions inspired by findings of cognitive psychology, yielding concepts like MF and MB.
In the recent and still ongoing \textit{Managed Forgetting} project, we especially focus on exploiting the yet untapped potential of more explicated user context and support measures inspired by Memory Inhibition.
We presented challenges that arise in the field of FIS, discussed how we tackled some of them and also gave insights on how we intend to solve the still open ones, which will be our focus in the remainder of the project.

\bibliographystyle{spmpsci}
\bibliography{paper}

\end{document}